\documentclass[a4paper,11pt]{article}
\pdfoutput=1 

\usepackage{jinstpub} 

\title{\boldmath Negative ion Time Projection Chamber operation with SF$_{6}$ at nearly atmospheric pressure}

\author[a,b1]{E. Baracchini,\note{Corresponding author.}}
\author[b]{G. Cavoto,}
\author[a]{G. Mazzitelli,}
\author[a]{F. Murtas,}
\author[b]{F. Renga,}
\author[a]{S. Tomassini,}

\affiliation[a]{INFN Laboratori Nazionali di Frascati, Via E. Fermi 40, 00044 Frascati, Italy.}
\affiliation[b]{INFN Sezione di Roma; Dipartimento di Fisica dell'Universita\` " La Sapienza", Piazzale
A. Moro, 00185 Roma, Italy.}
\emailAdd{elisabetta.baracchini@lnf.infn.it}

\abstract{We present measurements of drift velocities and mobilities of some innovative negative ion gas mixtures at nearly atmospheric pressure based on SF$_{6}$ as electronegative capture agent and of pure SF$_{6}$ at various pressures, performed with the NITEC detector. NITEC is a Time Projection Chamber with 5 cm drift distance readout by a GEMPix, a triple thin GEMs coupled to a Quad-Timepix chip, directly sensitive to the deposited charge on each of the 55 $\times$ 55 $\mu$m$^2$ pixel. Our results contribute to expanding the knowledge on the innovative use of SF$_{6}$ as negative ion gas and extend to triple thin GEMs the possibility of negative ion operation for the first time. Above all, our findings show the feasibility of negative ion operation with He:CF$_4$:SF$_{6}$ at 610 Torr, opening extremely interesting possibility for next generation directional Dark Matter detectors at 1 bar.}


\begin{document}
\maketitle
\flushbottom

\section{Introduction and Motivations}
\label{sec:intro}
Two of the most compelling questions that fundamental physics is facing today are the nature of Dark Matter (DM) and the properties of neutrinos. These topics involve the study of very rare processes and share the same challenging requirements on the simultaneous optimization of the active volume, energy resolution and background rejection capabilities of the detector. Extremely powerful tools in event discrimination are the topological signature and the directionality of the event. Although inherently challenging, gaseous TPCs potentially provide the best observables for a rare event search experiment, including the above features. They might even include anisotropic targets  with an additional 
directional sensitivity as suggested  in \cite{Cavoto:2016lqo,Capparelli:2014lua}.

A peculiar modification of conventional TPCs involves the addition to the gas mixture of a highly electronegative molecule, making it a Negative Ion TPC (NITPC). In this configuration, the primary electrons generated by the track during gas ionization are captured at very short distances (<= 100 um) by the electronegative molecules. Thanks to the TPC electric field, the resulting anions drift to the anode where they are ionized and normal electron avalanche occur. Since the anions act as image carriers instead of the electrons, their higher mass reduces longitudinal and transversal diffusion down to the thermal limit without need of magnetic fields. 

This idea was first proposed by Martoff in 2000~\cite{Martoff:2000wi} and was then successfully demonstrated with CS$_2$-based gas mixtures by DRIFT, a directional dark matter experiment~\cite{Alner:2005xp,Battat:2016xxe}. DRIFT presently employs a 30:10:1 Torr CS$_{2}$:CF$_{4}$:O$_{2}$ gas mixture, to exploit the fluorine content for Spin-Dependent (SD) DM interactions and the fiducialization provided by O$_{2}$. This last is a new decisive and peculiar feature of negative ions drift: after a small addition of oxygen, in fact, DRIFT recently observed the presence of secondary peaks in the time signals faster than the main one, and therefore identified as charge carriers with a mass different from CS$_{2}^{-}$~\cite{Snowden-Ifft:2014taa}. Given the different mobilities of anions of different masses, the difference between their time of arrival effectively provides a measurement of the position of the event in the drift direction. Since in most direct DM search experiments, a very annoying backgrounds are Radon Progeny Recoils (RPRs) coming from detector surfaces and mimicking a WIMP interactions, the possibility to establish the event position also in the drift direction allows to reject these events, opening attractive possibilities for next-generation detectors able to exploit such feature. Thanks to this technique, in fact, DRIFT managed to reach a background-free region of interest~\cite{Battat:2014van}.

Unfortunately, CS$_2$ has a combination of low flash point (-30$^{o}$ C), high vapor pressure (400 Torr) and low explosive mixture limit in air (1.3$\%$), which require great care in handling the material, especially when adding O$_{2}$. The utility of the NITPC would hence definitely be expanded by the identification of new negative ion mixtures and for this reason SF$_{6}$ has been recently proposed as an innovative, safer and more benign capture agent. Its non-toxicity, non-flammability, high vapor pressure, easiness to purify and recirculate make it easier to manage with respect to the DRIFT gas mixture, especially at an underground site. In particle detectors, SF$_6$ has been used as a quencher in Resistive Plate Chambers (RPCs), to suppress streamer formation (in avalanche RPCs) or reduce discharges (in streamer RPCs). This was possible thanks to its electron affinity, however initially regarded as too high for negative ion operation to allow the stripping of the electron that would initiate the avalanche in the amplification region. Recent years, however, saw a large development of new amplifying structures made from printed circuit-like substrates commonly denominated as Micro Pattern Gas Detectors (MPGDs), also in the DM search field~\cite{Battat:2016pap}. Thanks to the very small geometries obtainable, these devices can sustain very high electric fields and therefore provide extremely high gain, combined with high stability and granularity. 

The first experimental evidence for the possibility of using SF$_{6}$ as negative ion capture agent has been demonstrated by the pioneering studies by University of New Mexico~\cite{Phan:2016veo}, laying the groundwork for the future use of SF$_{6}$ in NITPC experiments. In this paper, they operate a single thick 400 $\mu$m GEM in the range 20-100 Torr of pure SF$_{6}$ and observed a gas gain between 2000-3000 on low energy 5.9 keV $^{55}$Fe x-ray clusters. They furthermore measured the mobilities of SF$_{5}^{-}$ and SF$_{6}^{-}$ up to high reduced fields and verified the diffusion properties, confirming the thermal behaviour at low $E/p$ reduced fields. Above all, they demonstrated the possibility of fiducialization along the drift direction through the use of minority carriers (SF$_{5}^{-}$), with results consistent with \cite{Snowden-Ifft:2014taa} with the DRIFT 30:10:1 Torr CS$_{2}$:CF$_{4}$:O$_{2}$ gas mixture. 

A NITPC approach with an He:SF$_{6}$ gas mixture (possibly at atmospheric pressure) with installation in multiple underground sites to minimize location systematics and improve sensitivity, has been recently proposed by the newly-formed collaboration CYGNUS for the next-generation directional DM experiment at the ton-scale. The gas choice is dictated by the exploitation of negative ion drift for reduced diffusion and SF$_{5}^{-}$ minority carriers for fiducialization, while at the same time extending the experiment sensitivity for spin-independent cross section to low WIMP mass thanks to the He.

Our work sets into this framework, in order to expand the study on SF$_{6}$ as capture agent for negative ion gas mixture, in the context of directional DM searches with gaseous TPCs.

\section{The NITEC Detector}
\label{sec:nitec}
\begin{figure}[hp]
\includegraphics[width=0.95\textwidth]{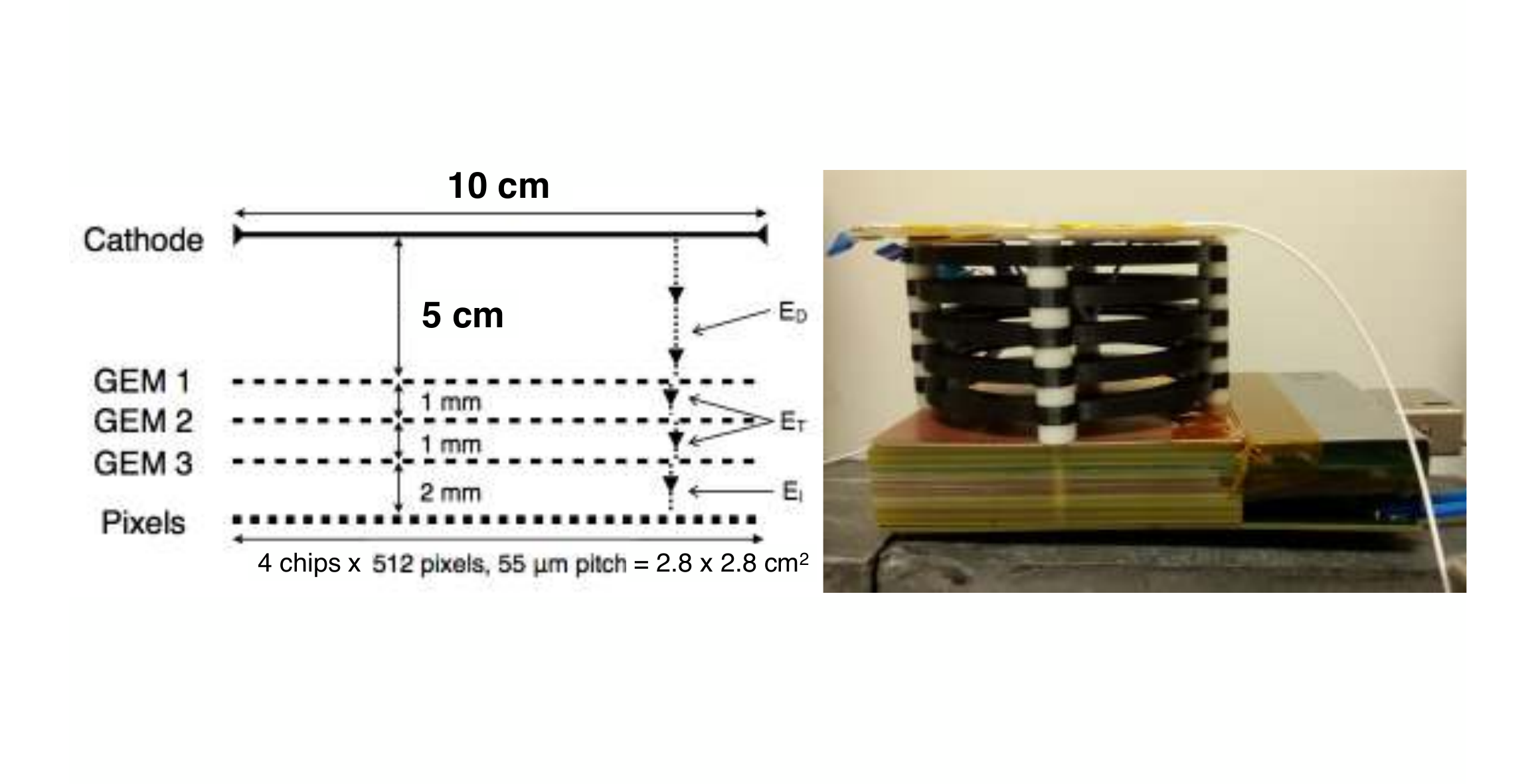}
\caption{\label{fig:nitec} NITEC detector: on the left a schematic showing the distances between the cathode, the GEMs and pixels planes, and on the right a picture showing the full detector, with the field cage polycarbonate field rings in black and the PCBs composing the GEMpix below.}
\end{figure}

The NITEC detector is composed by 5 cm long field cage with an inner diameter of 7.4 cm enclosed on one end by the cathode (a thin copper layer deposited on a PCB) and on the other by the GEMPix device. The 0.5 cm thick polycarbonate field rings, with 1 cm pitch, were manufactured with a 3D printer and 1 mm diameter silver wires were encased in to provide uniformity to the drift field. 1 GOhm resistors connecting the wires provide the partition of the HV through the field cage, that is provided by an external CAEN N1470 power supply. The GEMPix is obtained coupling three standard CERN thin GEMs with a 3 $\times$ 3 cm$^{2}$ active area to a Quad-Timepix ASIC with 262,144 pixels for readout, directly sensitive to the deposited charge (without any silicon sensors installed on top)~\cite{George:2015vfj}. A picture and a schematic of  NITEC is shown in Fig.~\ref{fig:nitec}.

Thin GEMs~\cite{Sauli:2016eeu} are MPGD composed by a very thin kapton insulating foil ($\sim$ 50 $\mu$m thick), electroplated with copper on both sides, where 70 $\mu$m width conical holes are etched with a chemical process. If a voltage is applied between the two side of the GEM foil, a very high electric field is generated in the holes up to 100 kV/cm. When an electron traverses the hole, avalanche multiplication takes place giving approximately 20 secondary electrons for each primary electron (with the exact value depending on gas density, gas mixture and applied electric field). The triple GEM configuration provides not only higher gas gain but also higher stability, with respect to single or double GEMs.  In the GEMPix, the GEM foils are held rigid by gluing them to a frame and then superimposing them on top of the pixels chip. The system is powered by a dedicated high voltage supply developed by Laboratori Nazionali di Frascati~\cite{Corradi:2007df}, which works as an active HV divider, with seven floating independent channels (for the 3 GEMs, plus the two transfer fields, the induction field and the drift field, not used in NITEC) isolated up to 5 kV to GND, each with a high sensitivity (nA) current meter. 

Timepix~\cite{Llopart:2007raz} is the second generation of the Medipix chip family, and is fabricated in the IBM 250 nm CMOS process, with a 256 $\times$ 256 matrix of 55 $\times$ 55 $\mu$m$^{2}$ square pixels. The processing electronics, including preamplifiers, discriminator threshold and pseudo-random counter (with 11818 counter depth), fit inside the footprint of the overlying pixel so that each one operates as an independent element. Since in our system the detecting medium is the gas, the silicon sensors are removed so that the pixels are directly sensitive to the deposited charge. The Timepix can be operated in counting mode (Medipix), but also in Time Of Arrival (TOA) or Time Over Threshold (TOT) mode,  thanks to a counting clock with an adjustable frequency up to 100 MHz which is propagated to every pixel, allowing to measure time or charge deposited. In TOA mode, the pixel counts the number of clk from when the preamplifiers exceeds the threshold to the end of the frame, while in TOT mode it counts the number of clk during which the signal is above threshold. This allows each pixel to act as a Wilkinson type ADC measuring the discharge time of the preamplifier. A very important and distinctive feature of the Timepix, not available in other chips used in this field (i.e. ATLAS chips in D3~\cite{Lewis:2014poa}) or in commonly employed DAQ electronics, is the adjustable sampling frequency, that can be lowered down to 50 kHz, allowing a gate window up to nearly 0.25 s. This is of paramount importance when working with negative ions drift, where typical sampling frequencies of 40-50 MHz are too fast since anions are 1000-10000 times slower than electrons. 

The Timepix is read out via the FITPix interface~\cite{Kraus:2011zza} (fully powered from an USB port), consisting of a field programmable gate array, a USB 2.0 interface chip, DAC, ADC and a circuit which generates bias voltage for the sensor (for Timepix chips with bumped bonded silicons). The main control system is placed in the FPGA circuit which fully controls the Timepix device and that can provide both hardware or software triggers. FITPix is controlled via the Pixelman software, developed for general Medpix2 chips control. A desktop computer controls both the HVGEM (via CANbus and a dedicated Labview program) and the FITPix interface (via USB and the Pixelman software).

\section{The Beam Test Facility and the Experimental Setup}
\label{sec:btf}
The Beam Test Facility (BTF) at Laboratori Nazionali di Frascati is a beam transfer line of the DA$\Phi$NE accelerator complex, that consists of a double ring electron-positron collider, a high current linear accelerator (LINAC), an intermediate damping ring and a system of 180 m transfer lines connecting the whole system~\cite{Ghigo:2003gy}. The LINAC provides an electron (positron) beam with energy up to 800 (550) MeV and a maximum current of 500 (100) mA. In order to tune beam intensity and energy, the BTF employs a copper target, a series of collimators and energy selectors through bending magnet and slits, that allows to achieve a momentum resolution on the selected beam of better than 1$\%$.  In this way, electrons of energy between 25 and 750 MeV with a repetition rate of 50 Hz with a 10 ns pulse duration are produced with a multiplicity between 1-10$^8$ particles per bunch~\cite{Mazzitelli:2006an}. After the energy selector, the beam is driven by a 12 m transfer line into the experimental hall by means of a focusing system of four quadrupoles.  In our setup, we choose to optimize the intensity of the beam rather than the dimension, given the short time available for the measurement. We used a 450 MeV electron beam with a multiplicity of about 500 particle per bunch. This resulted in a beam-spot dimension of about 1 mm in Y (the drift direction in the detector) and 2 mm in X , with about 2 mm divergence, as from the BTF beam monitoring.

\begin{figure}[hp]
\centering
\includegraphics[width=0.8\textwidth]{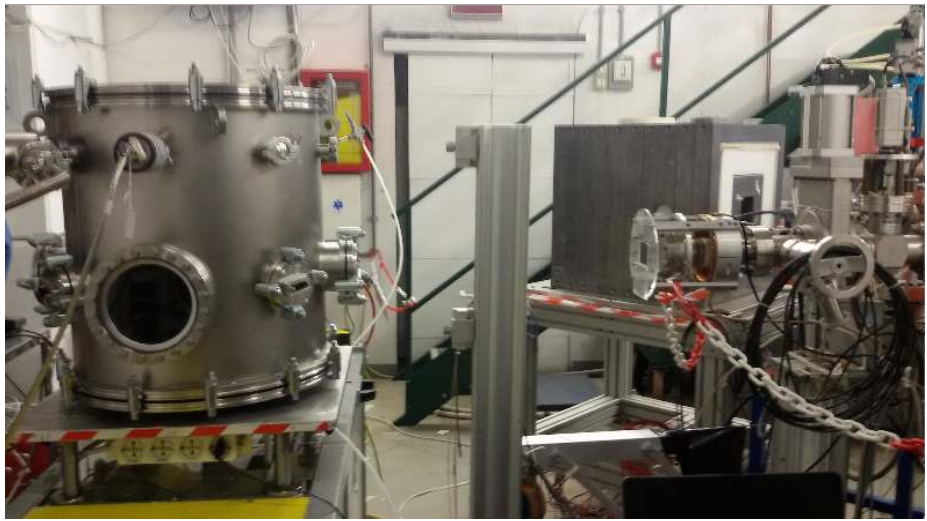}
\caption{\label{fig:setup} Experimental setup at the BTF, with the electron beam going from right to left into the vacuum vessel (cylindrical stainless steel container on the left).}
\end{figure}

In order to operate NITEC below atmospheric pressure, the detector was positioned inside a vacuum vessel, properly equipped with a 50 $\mu$m thick mylar window to allow the electron beam to enter the active volume with negligible loss of intensity and multiple scattering. The field cage and the Timepix were commonly grounded to the vessel from the inside, that in turn was connected to the experimental hall ground. The Fitpix and the HV for GEMs and field cage were operated from outside through custom feedthroughs. After sealing, the vessel was usually pumped down to about 0.1 Torr with a dry scroll vacuum pump and then back-filled with the chosen gas mixture to the operating pressure. We monitored the overall pressure inside the vessel during data taking with a capacitance barometer and repeated the above procedure any time the contamination reached about 1$\%$. The vessel was positioned on top of a movable table, that can be moved with micro-metric (less 10 $\mu$m uncertainty) precision in X and Y directions, at 110 cm distance from the exiting of the beam into the experimental hall. This allowed to precisely position the electron beam at different drift distances from the GEMPix readout and hence perform the mobility and drift velocities measurements here presented. A picture of the experimental setup showing the end of the beam line and the vacuum vessel is shown in Fig.~\ref{fig:setup}.

\section{Drift velocity and mobility measurements}
\label{sec:meas}
Several measurements in various conditions have shown that electron capture by the electronegative SF$_{6}$ molecule occurs very rapidly, with a mean free path of about a micron and the immediate product being SF$_{6}^{-*}$~\cite{Phan:2016veo, Christophorou}. The metastable SF$_{6}^{-*}$ de-excites into SF$_{6}$, SF$_{6}^{-}$ or SF$_{5}^{-}$ through auto-detachment, collisional stabilization or auto-dissociation respectively, with the relative abundances depend on the lifetime of SF$_{6}^{-*}$, the electron energy, gas pressure, temperature, and drift field. Auto-detachment is expected to be inconsequential at the operating conditions of our detector, as shown in \cite{Phan:2016veo}. Other negative ion species such as F$^{-}$ or SF$_{4}^{-}$ can be produced, but at much lower probabilities, due to the much lower cross section and higher electron energies required. Therefore, in our experiments we expect only two anion species, SF$_{6}^{-}$ and SF$_{5}^{-}$. 

The possibility to have two or more species of charge carriers is of paramount importance for event fiducialization in gas-based TPCs employed in DM and other rare event searches, as discussed in Sec.~\ref{sec:intro}. It has been recently shown, in fact, that since the mobility is inversely proportional to the mass of the drifting anion, the time difference between the peaks from different species effectively provides a measurement of the event position along the drift direction, allowing for precise three-dimensional fiducialization~\cite{Snowden-Ifft:2014taa}.  

The measurements in \cite{Phan:2016veo} show that SF$_{5}^{-}$ mobility is 6.9$\%$ higher than SF$_{6}^{-}$ at a $E/N$ of about 39 Td and 9.3 $\%$ larger at 158 Td. We performed our measurements in the range 3-30 Td. As discussed in Sec.~\ref{sec:btf} the ionization in our setup was provided by a 450 MeV electron beam, that we optimized for intensity rather than dimension. In our experimental setup, the electron beam entering the experimental hall with about 1-2 mm beam-spot and 2 mrad divergence gets further smeared by the air and other materials present before the gas active volume inside NITEC (a Timepix beam monitor, a 50 $\mu$m mylar window and the gas inside the vacuum vessel), for an additional total 1.9 mm contribution from multiple scattering. Given all of this and the 5 cm NITEC drift distance, we do not expect to be able to distinguish the minority carriers species in the data. Also ~\cite{Phan:2016veo},  in fact, was not able to separate SF$_{5}^{-}$ and SF$_{6}^{-}$ peaks over drift distances below 10 cm. For the same reason, we do not expect to be sensitive to the diffusion over such short distances and therefore to verify the thermal behaviour of the negative ion drift in our measurements. Nonetheless, we can easily determine anions drift velocities and mobilities using the time of arrival of the ionization cloud of the set of tracks in the beam, without losing any relevant information. 

For each drift velocity, we measured the time of arrival of the ionization cloud through the Timepix TOA at five different drift distances (1, 2, 3, 4 and 5 cm). For each gas mixture and pressure, we repeated the same measurement at the drift fields of 250 V/cm, 530 v/cm, 640 V/cm, 750 V/cm and 860 V/cm. The data acquisition was triggered externally through the Fitpix interface thanks to a signal provided by the BTF system monitor itself, so that start time of the event effectively represent the passage of the beam through the detector (after the proper delay from cable lines is subtracted). 
While giving an estimate of the effective gas gain used in these measurement is beyond the scope of this paper, comparing the measurement presented here with preliminary results with a$^{55}$Fe radioactive source, we can state we were operating in a range of gain between $\sim$ 2000-5000. 

Before any measurement, we perform chip equalization to adjust each pixel threshold as most homogeneously as possible. Since we implement this separately for each of the 4 chip of the Quad-Timepix, in our analysis we determine the drift velocity measured by each chip independently and then average the results. To do this, we perform a gaussian fit to the time of arrival distribution of the five drift distances in the range of $\pm$ 1 FWHM around the maximum and take the mean and sigma of the fit as the central value and error of the drift time. We then perform a linear fit of the drift times versus drift distances and extract the drift velocity. 

In order to verify the detector functioning and capability for such measurement, we started with a conventional electron carriers gas mixture as Ar:CO$_{2}$:CF$_{4}$ with a 45:15:40 ratio and measured a drift velocity of 2.06 $\pm$ 0.02 cm/$\mu$s in a 500 V/cm drift field, in nice agreement with already existing measurements~\cite{Alfonsi:2004gh}. Thanks to this result and confident of the proper TPC operation and drift velocity measurement capability, we moved to negative ion operation, first in pure SF$_{6}$, and then mixing it with Ar:CO$_{2}$ 70:30 and He:CF$_{4}$ 60:40 premixed gases. The choice of the gas mixtures has been dictated by the will to test the effect of both low ($\sim$10 Torr) and high ($\sim$ 100 Torr) SF$_{6}$ content and with the goal of demonstrating negative ion operation with He and SF$_{6}$ at nearly atmospheric pressure. 

\begin{figure}[htbp]
\centering
\includegraphics[width=0.4\textwidth]{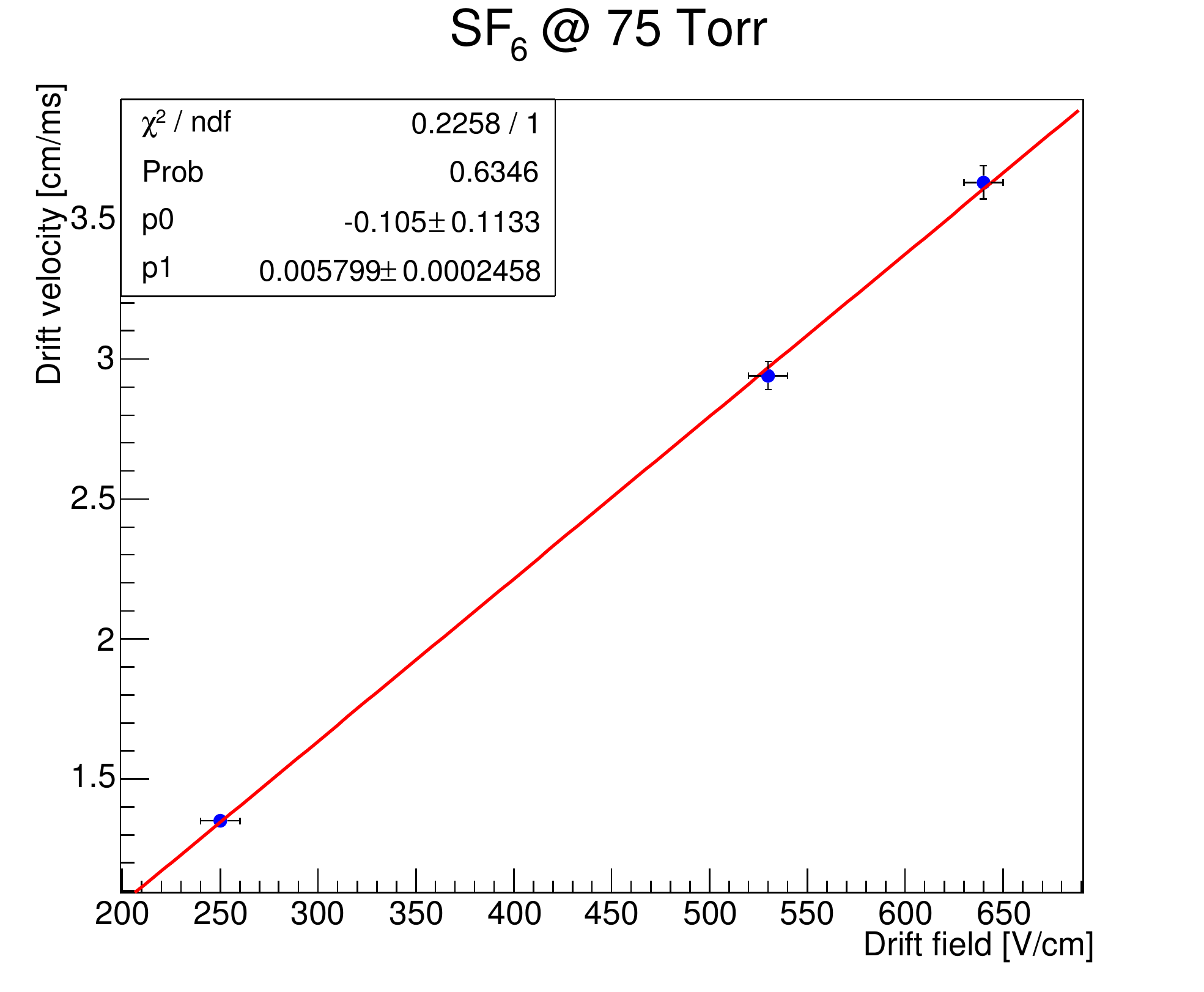}
\includegraphics[width=0.4\textwidth]{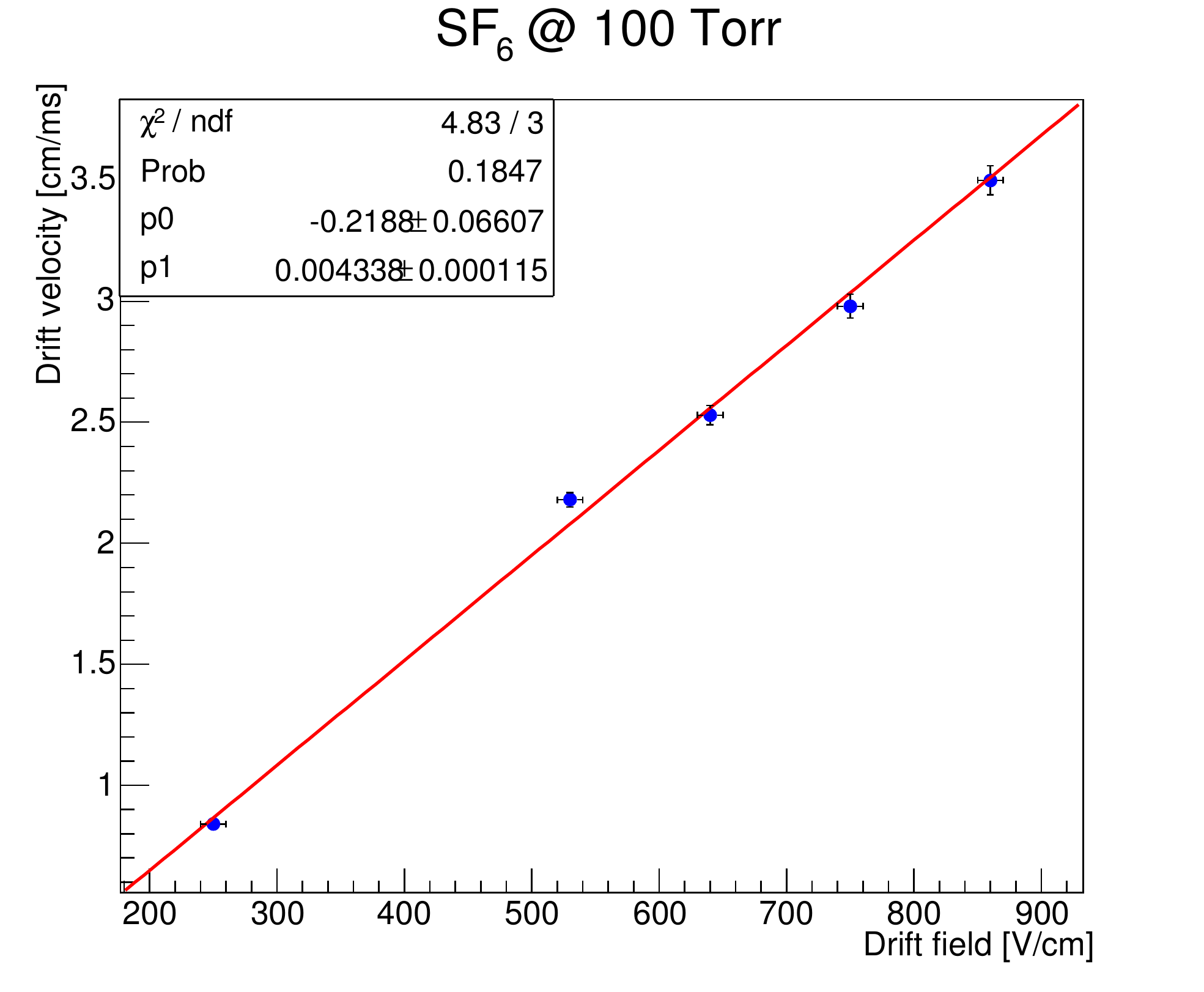}\\
\includegraphics[width=0.4\textwidth]{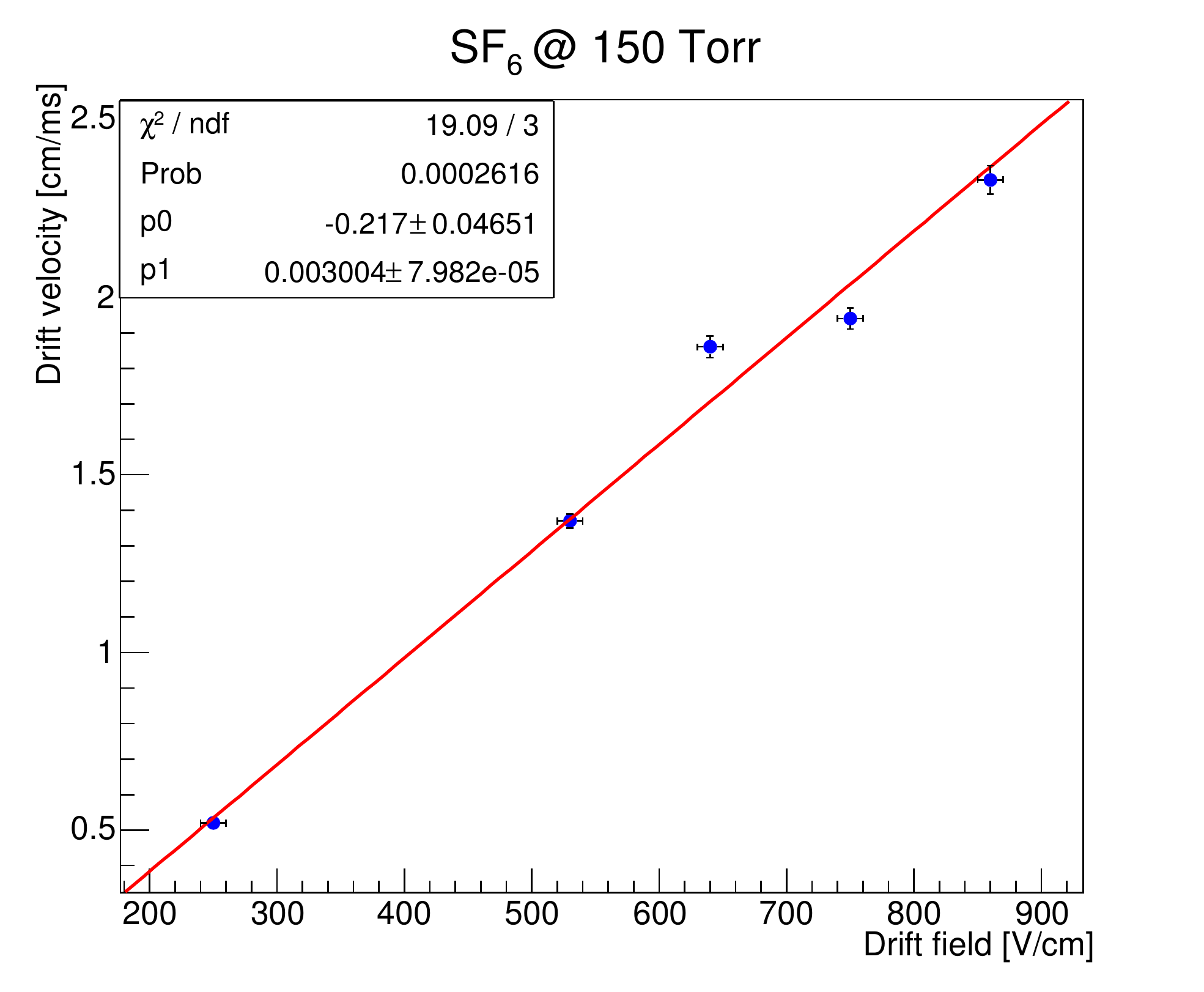}
\includegraphics[width=0.4\textwidth]{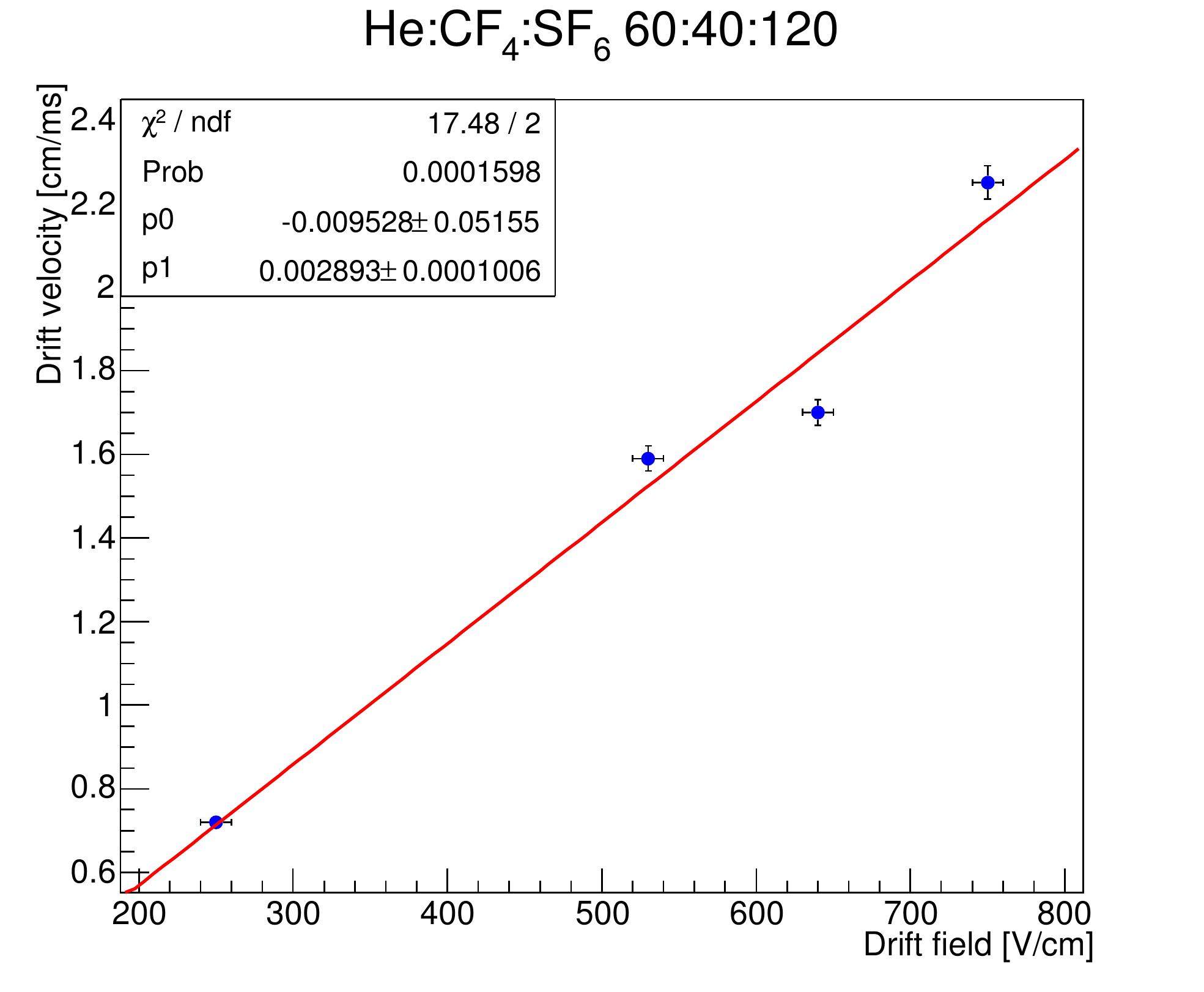}\\
\includegraphics[width=0.4\textwidth]{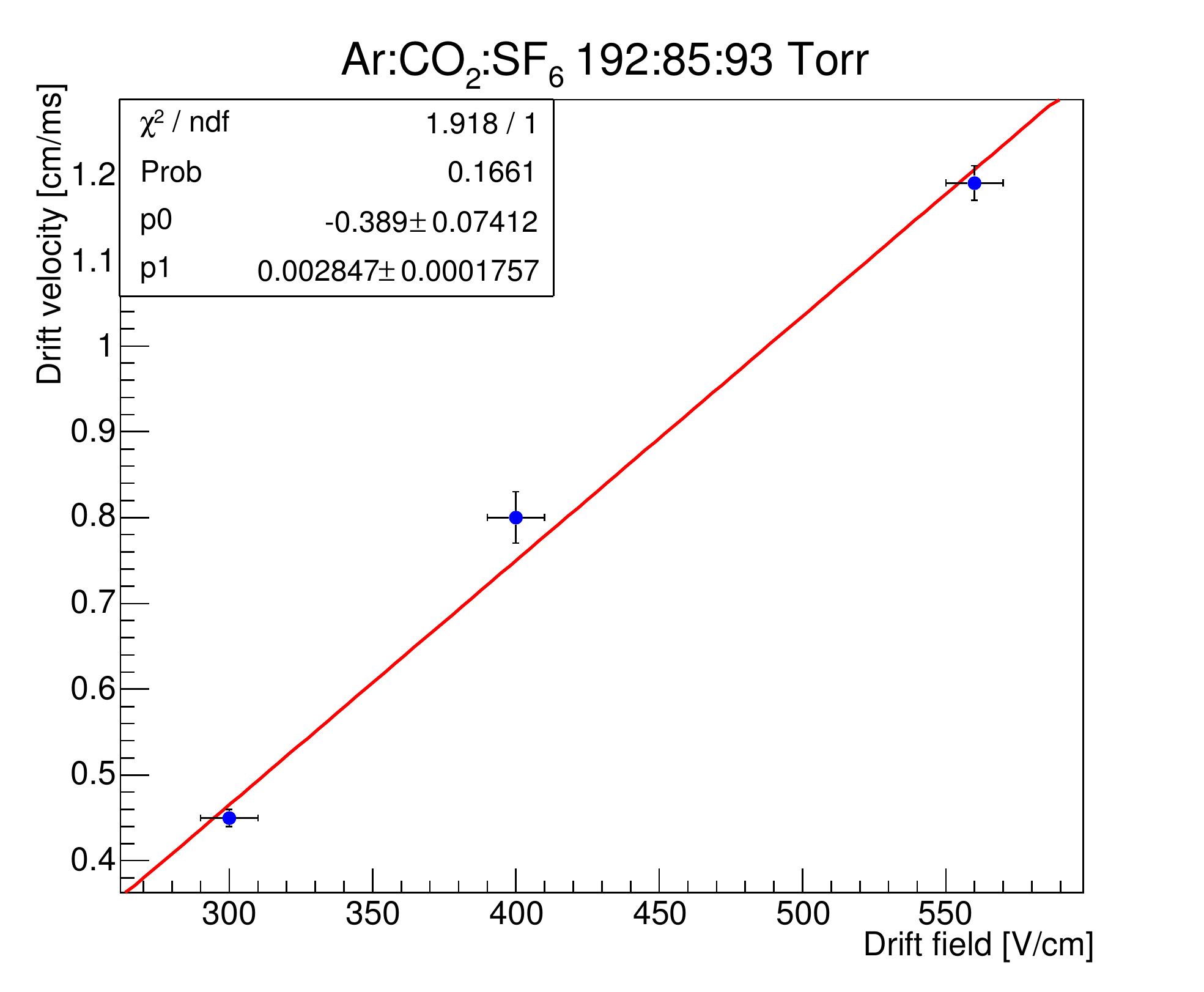}
\includegraphics[width=0.4\textwidth]{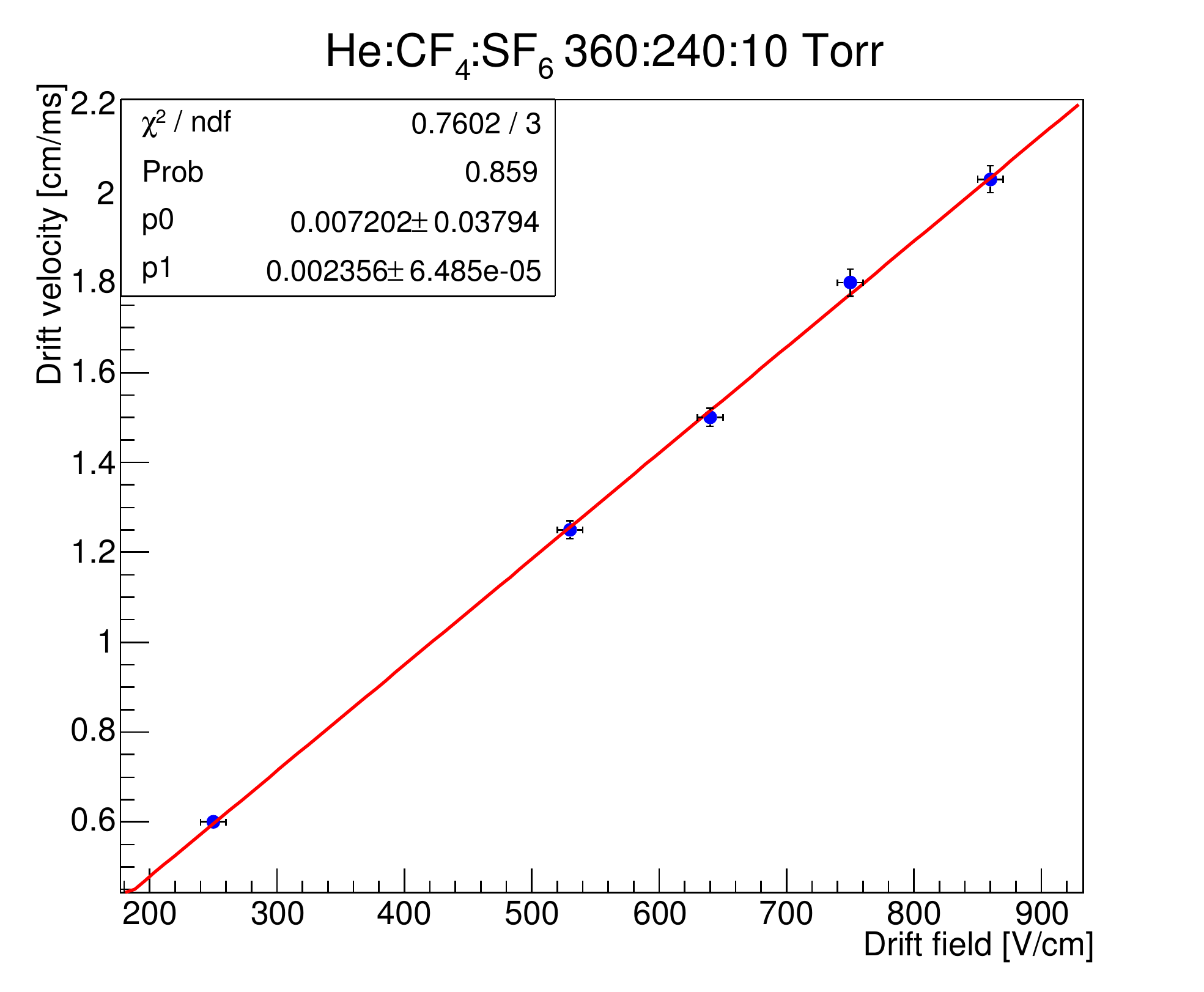}\\
\caption{\label{fig:drift} SF$_{6}^{-}$ drift velocities measured with NITEC. From top to bottom and from left to right, in pure SF$_{6}$ at 75 Torr, 100 Torr and 150 Torr, He:CF$_{4}$:SF$_{6}$ at 60:40:120 Torr, Ar:CO$_{2}$:SF$_{6}$ at 192:85:93 Torr and He:CF$_{4}$:SF$_{6}$ 360:240:10 Torr.}
\end{figure}

\begin{figure}[htbp]
\centering
\includegraphics[width=0.85\textwidth]{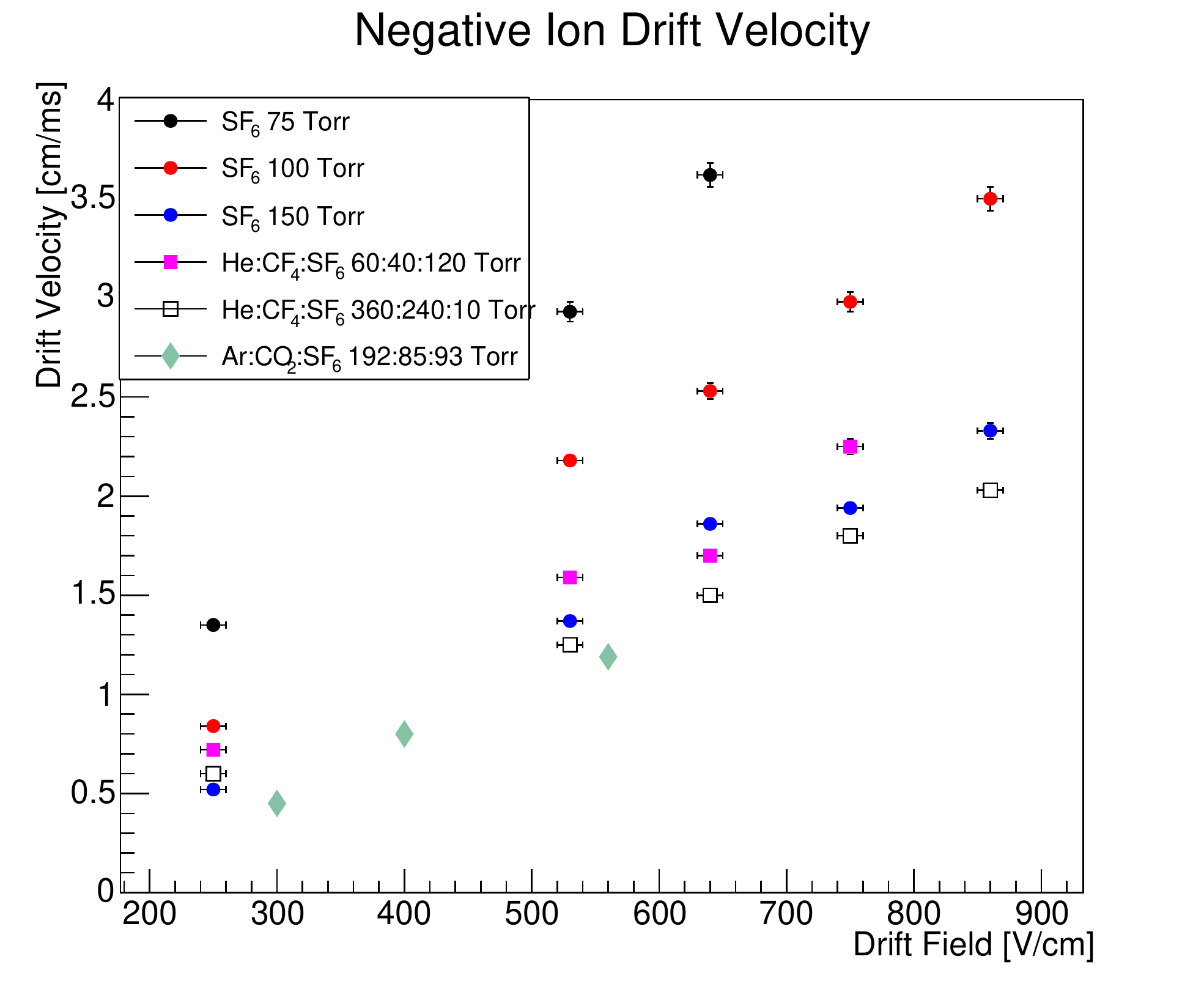}\\
\includegraphics[width=0.85\textwidth]{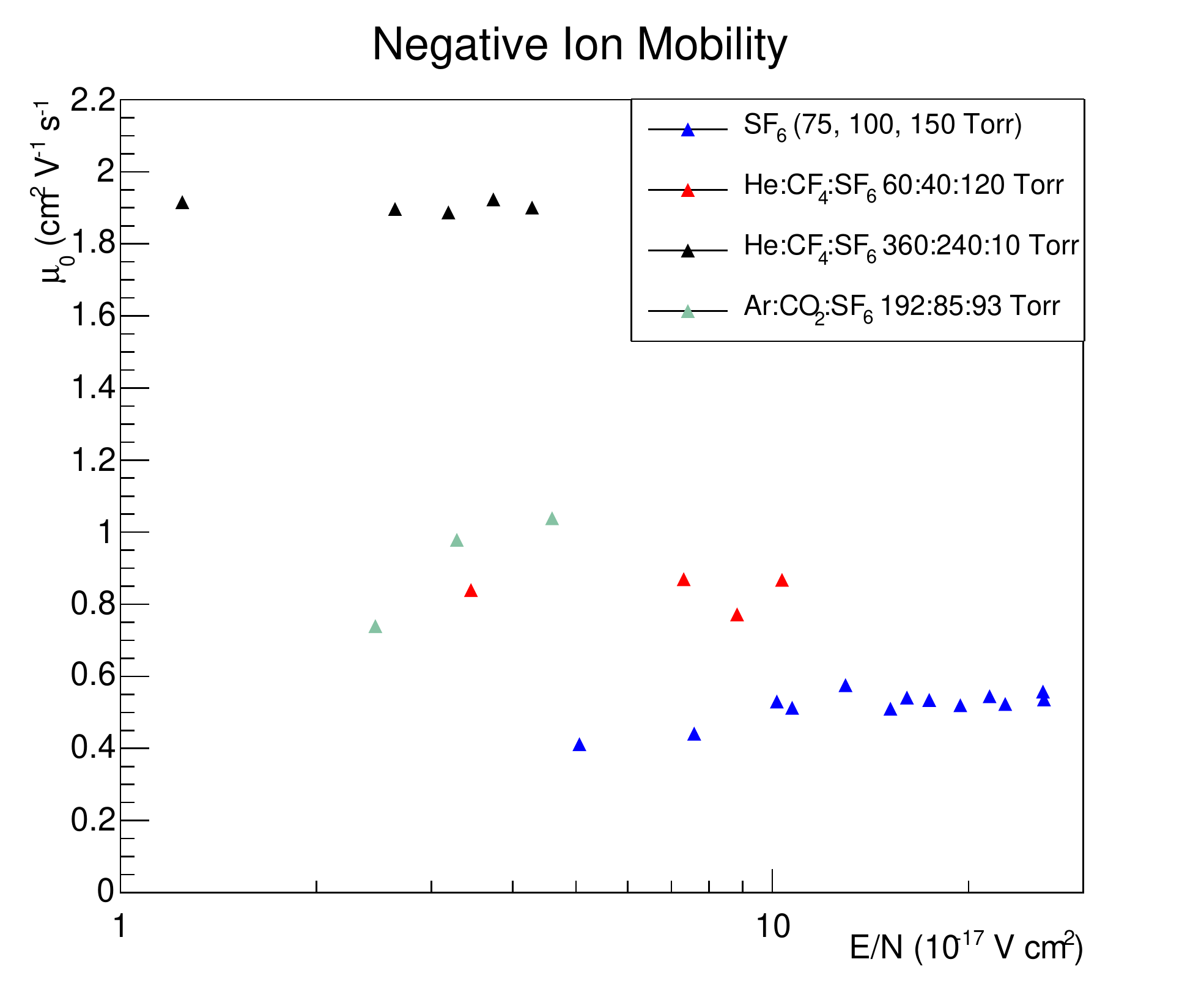}\\
\caption{\label{fig:mob} SF$_{6}^{-}$ drift velocities (top) and mobilities (bottom) in pure SF$_{6}$ (at 75 Torr, 100 Torr and 150 Torr), in Ar:CO$_{2}$:SF$_{6}$ at 192:85:93 Torr and He:CF$_{4}$:SF$_{6}$ at 60:40:120 Torr and 360:240:10 Torr. Given the $y$ axis scale, some errors unfortunately are not visible, especially in the bottom plot.}
\end{figure}

Fig.~\ref{fig:drift} shows the SF$_{6}^{-}$ drift velocities measured in the following gas mixtures, pressures and total HV applied to the triple GEMs: pure SF$_{6}$ at 75 Torr at 1140 V, 100 Torr and 150 Torr at 1240 V and 1440 V, Ar:CO$_{2}$:SF$_{6}$ at 192:85:93 Torr at 1480 V and He:CF$_{4}$:SF$_{6}$ at 60:40:120 Torr and 360:240:10 Torr at 1460 V and 1640 V respectively. Comparison of drift velocities and mobilities are shown in Fig.~\ref{fig:mob}. 

The SF$_{6}^{-}$ mobility we determined in pure SF$_{6}$ is consistent with the results from ~\cite{Phan:2016veo} in the region 10-30 Td of $E/N$ and extend the measurements below 10 Td. The drift velocities measured in the range of cm/ms demonstrate that all the other mixtures exhibit negative ion behaviour as well, with the highest SF$_{6}^{-}$ mobility obtained in the He:CF$_{4}$:SF$_{6}$ 360:240:10 Torr (as expected from the small SF$_{6}$ fraction), nearly 5 times higher than in pure SF$_{6}$. This is the first demonstration of SF$_{6}$ negative ion operation at nearly atmospheric pressure. The chosen mixture of He:CF$_{4}$:SF$_{6}$ 360:240:10 Torr, thanks to the low He density, permit nuclear recoil tracks lengths to remain long enough to be able to determine the direction even for low energy events. This, thanks also to the presence of He that increase the available mass and permit to extend sensitivity to the Dark Matter signal down to low 0.1-10 GeV WIMP masses, opens extremely interesting possibilities for next generation Dark Matter detectors for the development of a directional experiment at 1 bar.

\section{Conclusions}
\label{sec:conclusion}
NITPC are a peculiar modification of the TPC approach, whose reduced diffusion and possibility for fiducialization represent extremely interesting features for directional DM experiments. The recent discovery of the possibility to employ the much more safer and easy to handle SF$_6$ as capture agent opens striking new possibilities for next-generation detectors at the ton-scale, as recognized by the CYGNUS project. Along with these recent development in this sector, in this paper we studied SF$_{6}^{-}$ drift velocities and mobilities in various gas mixtures, thanks to the NITEC detector and a 450 MeV electron beam provided by the Beam Test Facility at Laboratori Nazionali di Frascati. Our results prove for the first time the possibility of negative ion operation with SF$_6$ at nearly atmospheric pressure (610 Torr). The chosen mixture of He:CF$_{4}$:SF$_{6}$ 360:240:10 Torr, thanks to the low He density that permit nuclear recoil tracks lengths to remain long enough to be able to determine the direction even for low energy events, is particularly interesting for directional Dark Matter searches. Thanks to the He content, in fact, the total experiment mass can be increased and sensitivity to the Dark Matter signal extended in the low 0.1-10 GeV WIMP mass region, while still maintaining directionality capability. This significant achievement can now open the doors for a realistic development of a negative ion TPC with He:SF$_{6}$ at 1 bar for directional Dark Matter searches, as foreseen by the CYGNUS collaboration.

\acknowledgments
This work has been supported by the European Union's Horizon 2020 research and innovation programme under the Marie Sklodowska-Curie grant agreement No. 657751.


\end{document}